\documentclass[prl,aps,reprint,twocolumn,nopacs,superscriptaddress]{revtex4}

\usepackage[utf8]{inputenc}
\usepackage[american,]{babel}
\usepackage[T1]{fontenc}
\usepackage[pdftex]{graphicx}  
\usepackage{graphicx, xcolor}
\usepackage{dcolumn}
\usepackage{bm}
\usepackage{amsmath,amsthm,amssymb}
\usepackage{hyperref}
\usepackage{xcolor}
\hypersetup{colorlinks,bookmarksopen,bookmarksnumbered,
citecolor=cyan,
linkcolor=cyan,
pdfstartview=false,
urlcolor=cyan}
\usepackage{graphicx}
\usepackage{braket}
\usepackage{wrapfig}
\usepackage{soul}
\usepackage{mathtools}
\usepackage{float}
\usepackage{natbib}

\renewcommand{\vec}{\boldsymbol}

\begin{document}

\title{Entanglement Transitions from Stochastic Resetting of Non-Hermitian Quasiparticles}

\author{Xhek Turkeshi}
\affiliation{JEIP, USR 3573 CNRS, Coll\`{e}ge de France, PSL Research University, 11 Place Marcelin Berthelot, 75321 Paris Cedex 05, France}
\author{Marcello Dalmonte}
\affiliation{SISSA, via Bonomea 265, 34136 Trieste, Italy}
\affiliation{ICTP, strada Costiera 11, 34151 Trieste, Italy}
\author{Rosario Fazio}
\affiliation{ICTP, strada Costiera 11, 34151 Trieste, Italy}
\affiliation{Dipartimento di Fisica, Universit\'a di Napoli "Federico II", Monte S. Angelo, I-80126 Napoli, Italy}
\author{Marco Schir\`o}
\affiliation{JEIP, USR 3573 CNRS, Coll\`{e}ge de France, PSL Research University, 11 Place Marcelin Berthelot, 75321 Paris Cedex 05, France}

\begin{abstract}
We put forward a phenomenological theory for entanglement dynamics in monitored quantum many-body systems with well-defined quasiparticles. Within this theory entanglement is carried by ballistically propagating non-Hermitian quasiparticles which are stochastically reset by the measurement protocol with rate given by their finite inverse lifetime. We write down a renewal equation for the statistics of the entanglement entropy and show that depending on the spectrum of quasiparticle decay rates different entanglement scaling can arise and even sharp entanglement phase transitions. When applied to a Quantum Ising chain where the transverse magnetization is measured by quantum jumps, our theory predicts a critical phase with logarithmic scaling of the entanglement, an area law phase and a continuous phase transition between them, with an effective central charge vanishing as a square root at the transition point. We compare these predictions with with exact numerical calculations on the same model and find an excellent agreement. \textbf{After the publication of the manuscript, we became aware of a numerical error in our implementation. We append the erratum to the end of this version.}
\end{abstract}

\date{\today}
\maketitle

Entanglement is a fundamental property of quantum mechanics, a key resource for emerging quantum technologies, and a powerful tool to characterize quantum phases of matter, in and out of equilibrium~\cite{amico2008entanglement,2009horodeckiquantum,laflorencie2016quantum}. For ground-state critical systems in one dimension it is by now understood that entanglement entropy scales logarithmically with system size, with a prefactor given by the central charge of the associated conformal field theory~\cite{calabrese2004entanglement}. Gapped systems on the other hand satisfy an area law~\cite{hastings2007area}. 
For translation-invariant systems evolving under unitary dynamics the entanglement growth is generally expected to be linear in time and to lead to a volume law scaling, both for integrable systems~\cite{calabrese2005evolution,alba2017entanglement} as well as for ergodic quantum many-body systems~\cite{kim2013ballistic,znidaric2020entanglement}. For random unitary circuits it was shown that fluctuations around this linear growth contain important information on the nature of the many-body dynamics and an effective mapping to  models of stochastic interface growth was discovered~\cite{nahum2017quantum}. Furthermore, the slow, logarithmic, growth of entanglement entropy in highly excited disordered quantum many-body systems is considered among the key signatures of Many-Body Localizaton~\cite{abanin2019colloquium}. 

Recently, a new class of phase transitions characterized by a qualitative change of the entanglement properties has been identified for quantum many-body systems under the effect of stochastic measurements~\cite{li2018quantum,li2019measurementdriven,skinner2019measurementinduced,chan2019unitaryprojective}. For hybrid quantum circuits with random unitary gates and projective measurements this protocol leads to an entanglement transition between an error correcting/volume law phase and a Zeno phase with area-law scaling~\cite{li2018quantum,li2019measurementdriven,skinner2019measurementinduced,nahum2021measurement,bao2021theory,gullans2019dynamical,gullans2019scalable,medina2021entanglement,sierant2021dissipative,choi2020quantum,hashizume2021measurementinduced}. Several aspects of this transition have been discussed, including its critical properties~\cite{vasseur2019entanglement,jian2019measurementinduced,piqueres2020meanfield,ware2021measurements,zabalo2019critical,zabalo2021operator,agrawal2021entanglement,li2021statistical,sierant2021universal}, the role of symmetries~\cite{bao2021symmetry} and dimensionality~\cite{lunt2021measurementinduced,turkeshi2020measurementinduced,block2021measurementinduced,sharma2021entanglement}. Recently an experimental implementation with trapped ions has been achieved~\cite{noel2021observation}.
\begin{figure}[t!]
	\includegraphics[width=\columnwidth]{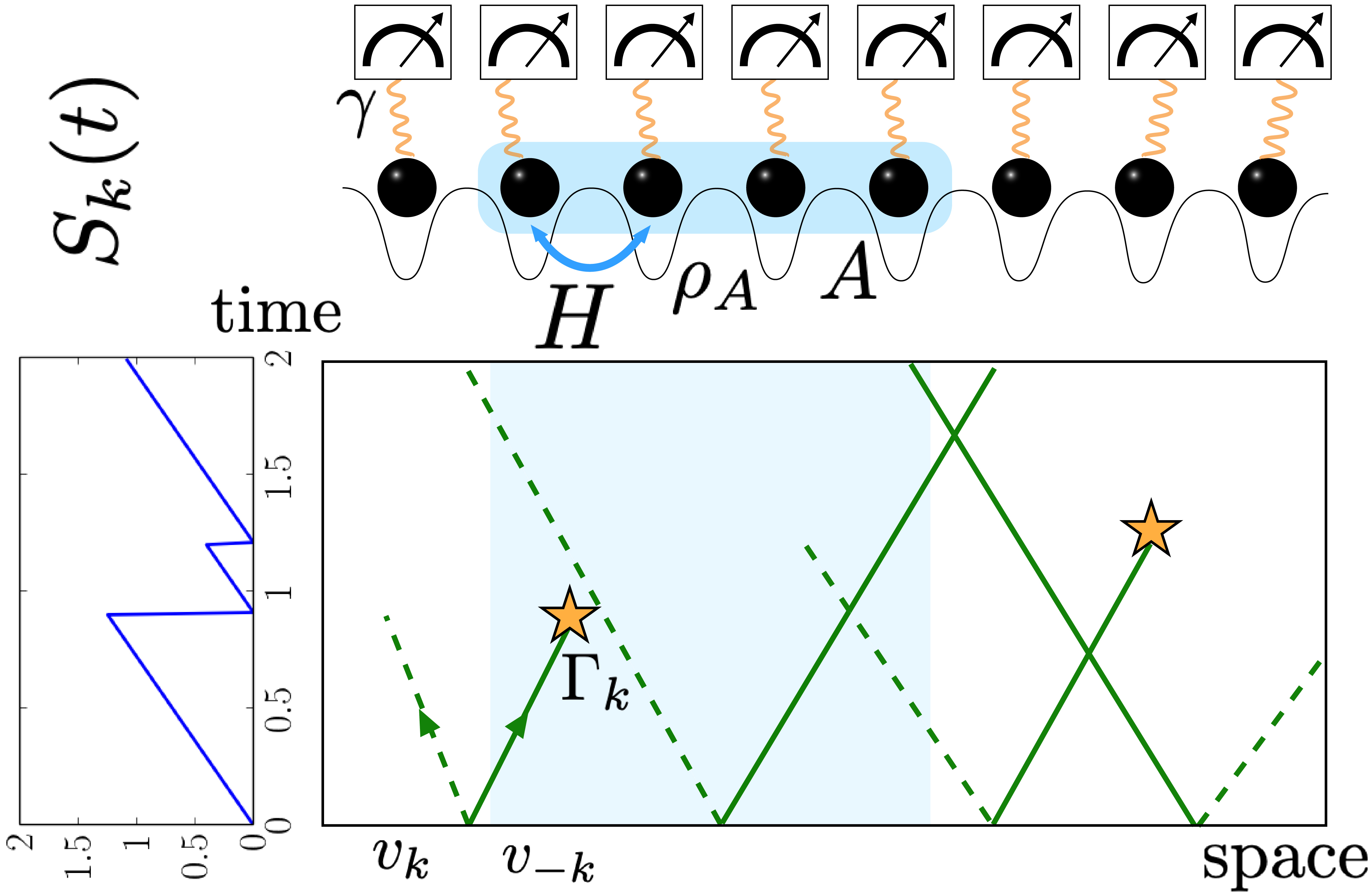}
	\caption{\label{fig:cartoon} (Top) A quantum many-body system evolves under the action of its own Hamiltonian and local quantum jump measurements with rate $\gamma$. Its entanglement entropy of a partition $A$ is obtained from the reduced density matrix $\rho_A$. (Bottom) Cartoon of the phenomenological picture for the entanglement dynamics where non-Hermitian quasiparticles with velocity $v_k$ are stochastically reset by the measurement with a rate given by their finite inverse lifetime $\Gamma_k$.}
\end{figure}
When the unitary dynamics is generated by a Hamiltonian one can expect a richer pattern of entanglement properties, which has just started to be unveiled. In the case of free fermions it was shown that the volume law is unstable for any value of the measurement strength to a subextensive entanglement content~\cite{cao2019entanglement,fidkowski2021howdynamicalquantum,coppola2021growth,muller2021measurementinduced,
botzung2021engineered,tang2021quantum}. Nonetheless an entanglement transition between a logarithmic and area law phase at a critical measurement strength was found in several translationally invariant free-fermionic models under different measurement protocols corresponding to the continuous monitoring limit~\cite{alberton2020entanglement,turkeshi2021measurementinduced_1,buchold2021effective,muller2021measurementinduced}. These results raise the question of whether a simple picture for entanglement transitions in these settings might exist, generalizing the quasi-particle picture available for integrable systems under unitary evolution~\cite{calabrese2005evolution,alba2017entanglement}. A first attempt in this direction lead to the collapsed quasiparticle ansatz of Ref.~\cite{cao2019entanglement}, predicting an area law scaling for any finite measurement rate, that appears at odds with the above results for monitored free-fermions.

In this Letter we put forward a phenomenological theory for entanglement dynamics in continuously monitored quantum many-body systems with well-defined quasi-particles and we test it against exact numerical simulations on the quantum Ising chain evolving under quantum jumps.
We assume that due to the measurement process the ballistically propagating quasiparticles, responsible for entanglement growth in the unitary case, acquire now a finite lifetime given by the associated non-Hermitian Hamiltonian~\cite{Wiseman2009}. We then describe the entanglement dynamics as a stochastic resetting process~\cite{evans2011diffusion,evans2020stochastic} , where those non-Hermitian quasiparticles are randomly reset with a rate given by their inverse lifetime. We write down a renewal equation for the entanglement statistics and compute its average value, from which we obtain that entanglement growth is directly connected to the spectrum of decay rates of those non-Hermitian quasiparticles. We show that our phenomenological picture captures all the key features of the monitored Ising chain, including the critical logarithmic entanglement phase, related to the existence of slowly decaying quasiparticles, its transition into an area law phase and an effective central charge vanishing continuously at the transition.
 
\emph{Unitary Dynamics and Measurement Protocol ---}  We consider a quantum many-body system evolving under the combined effect of a unitary evolution generated by a Hamiltonian $H$ and a measurement apparatus acting occasionally but abruptly on the quantum state to measure a local operator $L_i$, leading to Quantum Jump (\textbf{QJ}) trajectories described by the following stochastic Schr\"odinger equation~\cite{Dalibard1992,dum1992montecarlo,carmichael1993quantum,Plenio1998,Daley2014,Wiseman2009}
\begin{align}\label{eq:qjump}
d|\psi(\vec{\mathcal{N}}_t )\rangle &= \left[-iH-\frac{\gamma}{2}\sum_i \left(L^\dagger_i L_i-\langle L^\dagger_i L_i\rangle_t \right)\right]dt|\psi(\vec{\mathcal{N}}_t)\rangle \nonumber \\
&\quad + \sum_i\left(\frac{L_i}{\sqrt{\langle L^\dagger_i L_i\rangle_t}}-1\right) d\mathcal{N}_t^i|\psi(\vec{\mathcal{N}}_t)\rangle,
\end{align}
where ${\langle \circ \rangle_t\equiv \langle \psi(\vec{\mathcal{N}}_t) |\circ |\psi(\vec{\mathcal{N}}_t)\rangle}$, $\gamma$ is the measurement rate, and ${\vec{\mathcal{N}}_t=\{\mathcal{N}_{i,t}\}}$ are Poisson processes  ${d\mathcal{N}_{i,t}=0,1}$, statistically independent ${d\mathcal{N}_{i,t} d\mathcal{N}_{j',t} = \delta_{i,j} dN_{i,t}}$, and with average value $\overline{d\mathcal{N}_{i,t}} = \gamma dt \langle L^\dagger_i L_i\rangle_t$. We assume the unitary evolution, the first term in~\eqref{eq:qjump}, to be described by a Hamiltonian containing well-defined quasiparticle degrees of freedom and the operator being measured $L_i$ to be local in these quasiparticles -- see below for a specific example.

 In the following we will be mainly interested in the (von Neumann) entanglement entropy, defined as ~\cite{calabrese2004entanglement,amico2008entanglement,2009horodeckiquantum}
\begin{align}
	S(\vec{\mathcal{N}}_t )  = -\mathrm{tr}_A\left[ \rho_A(\vec{\mathcal{N}}_t)\ln \rho_A(\vec{\mathcal{N}}_t)\right]\;,
	\label{eq:5v0}
\end{align}
where we have introduced a partition $A\cup B$ and the reduced density matrix $\rho_A(\vec{\mathcal{N}}_t) = \mathrm{tr}_B |\psi(\vec{\mathcal{N}}_t)\rangle\langle\psi(\vec{\mathcal{N}}_t)|$. In particular we will focus on the conditional average entanglement entropy, given by $\overline{S}_t = \int \mathcal{D}\vec{\mathcal{N}}_t P(\vec{\mathcal{N}}_t) S(\vec{\mathcal{N}}_t)$.

\textit{Stochastic Resetting of Non-Hermitian Quasiparticles ---} We now present our phenomenological theory which takes the form of an effective stochastic process for the entanglement dynamics. In absence of any measurement, it is well known that entanglement is carried by pairs of quasiparticles with momenta $k,-k$ moving along the light-cone with velocity $v_k$. The measurement protocol in ~\eqref{eq:qjump} gives rise to two effects, namely a (i) non-unitary evolution described by a non-Hermitian Hamiltonian~\cite{Wiseman2009} $H_{\rm eff}=H-i\frac{\gamma}{2}\sum_i L^{\dagger}_iL_i$ and (ii) stochastic quantum jumps  which tend to suppress entanglement. We assume that the non-Hermitian Hamiltonian can be still written in terms of quasiparticles with momentum $k$   
which propagate ballistically but acquire a finite decay rate $\Gamma_k$ given by the imaginary part of their complex energy $\Lambda_k=E_k+i\Gamma_k$. Drawing from the literature on stochastic resetting~\cite{evans2011diffusion,gupta2014fluctuating,majumdar2015dynamical,evans2020stochastic,magoni2020ising} we postulate that the combined effect of measurements is to randomly reset these non-Hermitian quasiparticles, according to a Poisson process with a rate that we identify with their inverse lifetime $\Gamma_k$ and that therefore strongly depends on the competition between unitary dynamics and local measurements.

The key quantity in our theory is therefore the probability $P_k(s,t)$ of having at time $t$ a contribution $s$ to the entanglement entropy due to quasiparticles at momenta $k,-k$, propagating with velocity $v_k$ and being reset with a rate $\Gamma_k$. An infinite number of resetting events contribute to this probability, which can be resummed into a renewal equation of the form~\cite{SM,fagotti2011entanglement,turkeshi2020negativity,turkeshi2020entanglement}
\begin{align}\label{eqn:renewal}
P_k(s,t)=e^{-\Gamma_k t}p_k^0(s,t)+{\int\limits_0^t} d\tau \Gamma_k e^{-\Gamma_k\tau}P_k(s,t-\tau)
\end{align}
with $p^0_k(s,t)=\delta(s-s^*_k(t))$ with $s^*_k(t)=s_0\mbox{min}\left(2\vert v_k\vert t,\ell\right)$ describing the ballistic contribution of the isolated system with $s_0$ the thermodynamic entropy density~\cite{fagotti2008evolution,alba2017entanglement,calabrese2020entanglement}. We further assume the initial state of the dynamics to be completely uncorrelated and that the resetting process restarts the entanglement trajectory $s^*_k(t)$ from its product state value.

The two terms in~\eqref{eqn:renewal} have a clear meaning as renewal processes, where the first term describes a situation where the no resetting has taken place up to time $t$, whose probability is $e^{-\Gamma_k t}$ and the entanglement statistics is given by the unitary evolution, while the second describes the situation in which the first resetting event occurs at time $\tau$ and the following evolution during $t-\tau$ accounts for all possible resetting events. An equivalent formulation in terms of the last resetting time can be also obtained~\cite{SM}. To solve the renewal equation it is convenient to introduce the characteristic function $F_k(q,t)$ defined as
\begin{align}
F_k(q,t)=\int ds e^{-iqs}P_k(s,t)\;.
\end{align}
A simple calculation gives~\cite{SM} 
\begin{align}\label{eqn:Fk}
F_k(q,t)=\frac{\Gamma_k+iq B_k(t) e^{-iq \tilde{B}_k \mbox{min}(t,t^*)} }{\Gamma_k+iq\tilde{B}_k}\;,
\end{align}
with $\tilde{B}_k=2s_0\vert v_k\vert$, $B_k(t)=\tilde{B}_k\,e^{-\Gamma_k\mbox{min}(t,t^*)}$ and $t^*=\ell/2\vert v_k\vert$.  
Let us now discuss the consequences of this ansatz for the average entanglement entropy, which can be written after simple manipulations as~\cite{SM}
\begin{equation}\label{eqn:Save_dyn}
\overline{S}_t(\ell)=
\int \frac{dk}{2\pi}
\frac{2\vert v_k\vert}{\Gamma_k}
\left[1-e^{-\Gamma_k \mbox{min}(t,t^*)}
\right]\;.
\end{equation}
Taking the long-time limit we obtain an expression for the stationary state entanglement entropy which reads
\begin{equation}\label{eqn:Save_ss}
\overline{S}_{\infty}(\ell)=\,\int\frac{dk}{2\pi} \frac{2\vert v_k\vert }{\Gamma_k}\left(1-e^{-\ell\Gamma_k/2\vert v_k\vert }\right)\;.
\end{equation}
We see that for a resetting rate given by the \emph{bare} measurement rate, i.e. $\Gamma_k\equiv \gamma$,  our theory reduces to the collapsed quasi-particle ansatz of Ref.~\cite{cao2019entanglement} leading, for any measurement rate, to an exponential saturation of the entanglement in time and a stationary state with area law. As we are going to show, this scenario can dramatically change if the decay rate $\Gamma_k$ vanishes for certain $k$ points, leading to a gapless dispersion of decay modes. In this case, the integral in~\eqref{eqn:Save_dyn} might diverge with either time or system size and different scaling of the entanglement entropy can be obtained depending on the nature of the decay rate.

\emph{Application: Quantum Ising chain with Quantum Jumps --- }  To test our quasiparticle picture we consider a quantum Ising chain, where the transverse magnetization is measured through a QJ protocol. Specifically, we take in \eqref{eq:qjump} the Hamiltonian of the form
\begin{align}
	H = -J\sum_{i=1}^{L-1}\sigma^x_i \sigma^x_{i+1}-h\sum_i\sigma^z_i
\end{align}
with open boundary conditions, where $\sigma^\alpha$ are Pauli matrices. We note that with respect to Ref.~\cite{turkeshi2021measurementinduced_1} here we include also a transverse field $h>0$ in the unitary part of the evolution. For what concerns the measuring process we choose the local operator in \eqref{eq:qjump} as $L_i=(1+\sigma^z_i)/2$ and we start at $t=0$ from the ground-state of $H$ with $J=1$ and $h=h_0\gg 1$~\cite{SM}. To compute the entanglement entropy we take a cut of length $\ell=L/4$. 
\begin{figure}[t!]
	\includegraphics[width=\columnwidth]{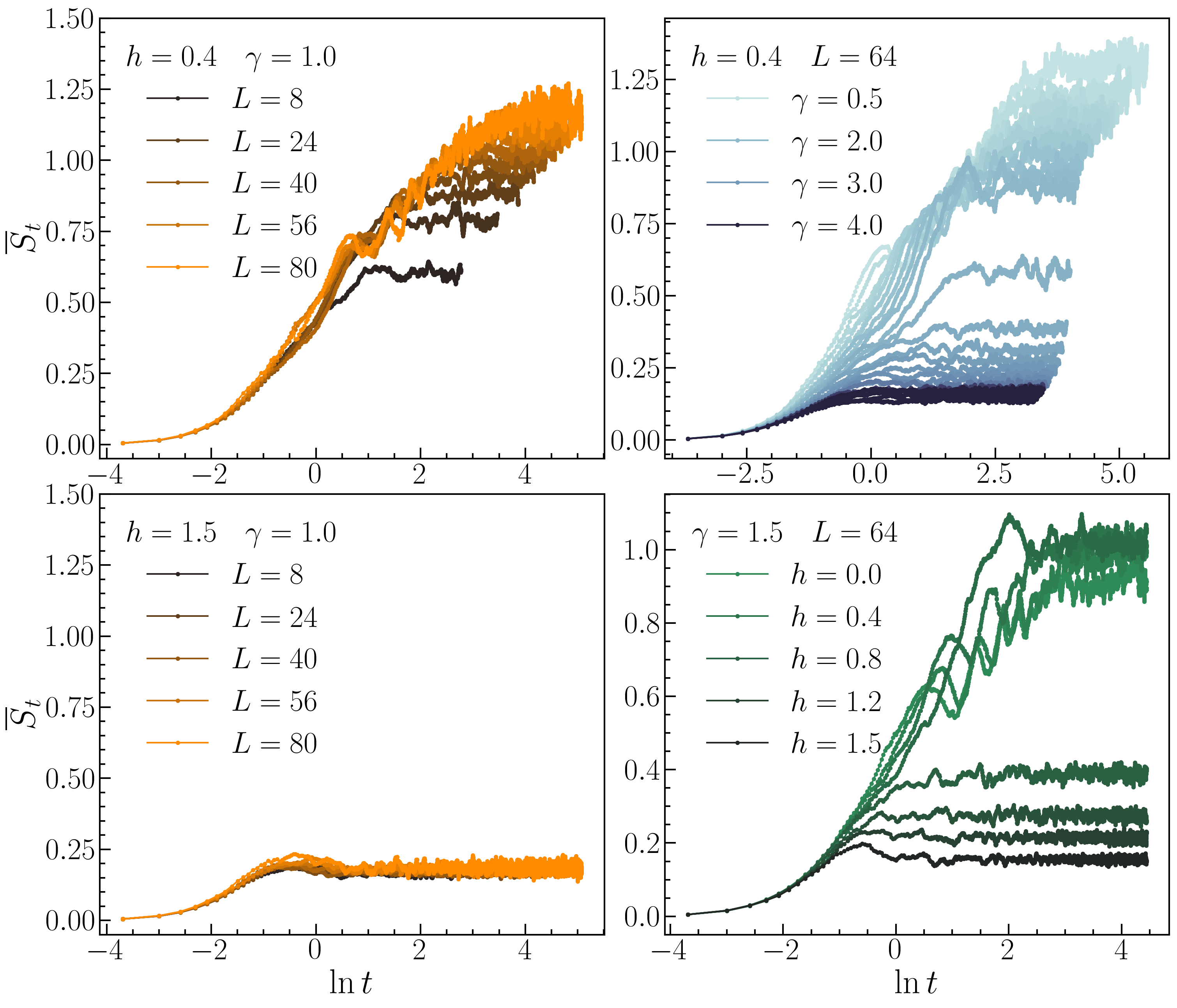}
	\caption{\label{fig:dynamics} Average entanglement entropy dynamics for different values of transverse field $h$ and measuring strength $\gamma$, obtained from the QJ protocol.  Top panels: logarithmic growth of the entanglement for $\gamma<\gamma_c$ and $h<h_c$ (left), which turns into an exponential approach to a stationary value upon increasing $\gamma$ above $\gamma_c(h)$ (right). Bottom panels: Entanglement dynamics for $h>h_c$ shows a weak dependence from system size (left), while upon decreasing $h$  a transition into a logarithmic growth phase emerges.}
\end{figure}
We first discuss the predictions from our phenomenological theory. To this extent we consider the non-Hermitian Hamiltonian associated with the QJ protocol which in the case under consideration takes the form of a quantum Ising chain with complex transverse field~\cite{hickey2013timeintegrated,lee2014heralded,biella2021manybody}, $H_{\rm eff}=H-i\gamma/4\sum_i\sigma^z_i$. This model can be diagonalized exactly in terms of non-Hermitian quasiparticles with complex spectrum $\Lambda_k$
\begin{align}\label{eqn:Lambdak}
\Lambda_k = \sqrt{\varepsilon_k^2-\gamma^2/4+2i\gamma\left(h-\cos k\right)}\equiv E_k+i\Gamma_k
\end{align}
and finite lifetime $\Gamma_k$. In \eqref{eqn:Lambdak} $\varepsilon_k=\sqrt{4\left(h^2+1-2h\cos k\right)}$ is the dispersion relation of the Hermitian Ising chain, featuring a quantum critical point at $h=h_c=1$ for ordering at $k=0$. 
\begin{figure*}[t!]
	\includegraphics[width=\textwidth]{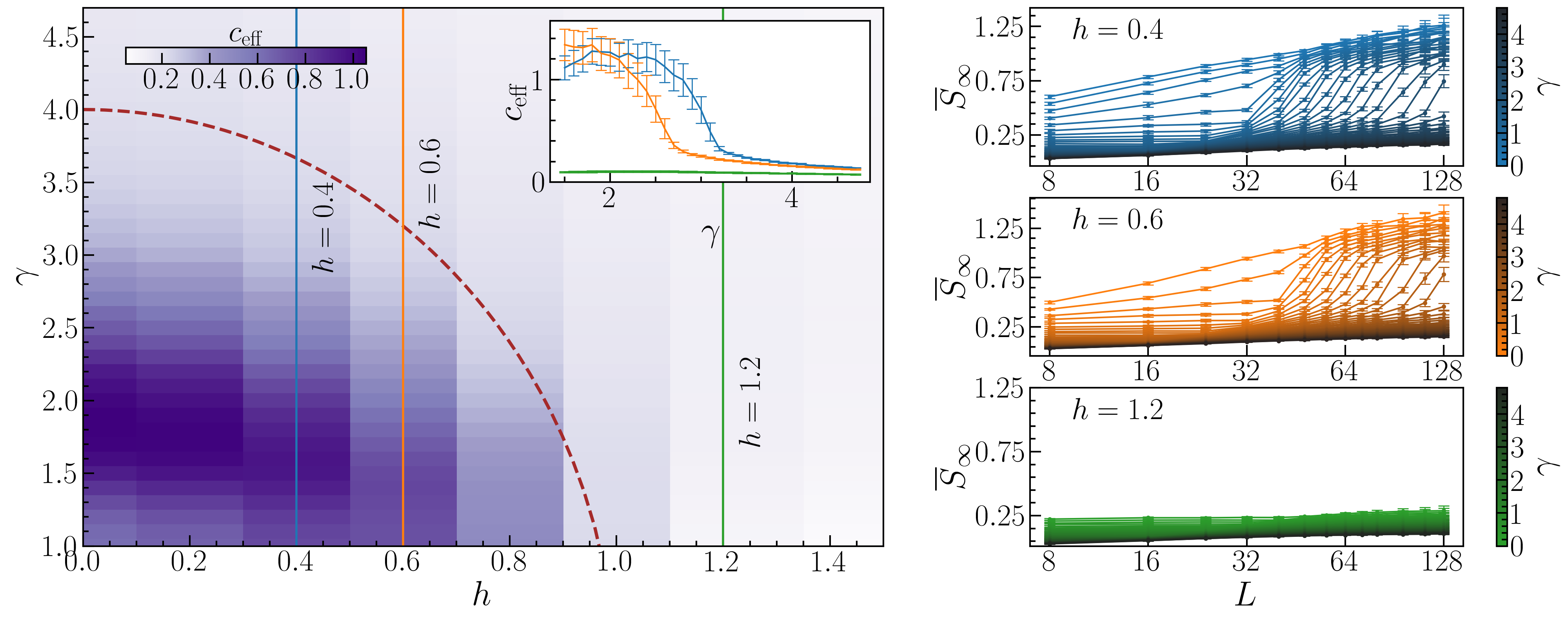}
	\caption{\label{fig:steadystate} Steady-state entanglement transition in the quantum Ising chain with QJ measurements. Left-panel: Phase diagram in the $(\gamma,h)$ plane featuring a critical logarithmic growth phase for $\gamma<\gamma_c(h)$ and an area-law (Zeno) phase elsewhere. Right-panels: Stationary entanglement versus system size for different values of $\gamma,h$ showing the logarithmic growth phase and the area-law. Inset: Effective central charge $c_{\rm eff}(\gamma)$ for three values of $h$ (see vertical cuts on the phase diagram).}
\end{figure*}
It is easy to see~\cite{SM} that the non-hermitian quasiparticle spectrum in \eqref{eqn:Lambdak} features a spectral (subradiant) transition~\cite{biella2021manybody,gopalakrishnan2020entanglement}, at a critical value of the measurement strength where the properties of the decay rates $\Gamma_k$ change qualitatively. As we are going to show now within our phenomenological theory, this directly affects the scaling of the entanglement entropy and leads to a sharp entanglement transition.  To present our results it is useful to distinguish two cases, namely $ h<h_c$ and $ h>h_c$. In the first case, we have that for $\gamma<\gamma_c(h)=4\sqrt{1-h^2}$ the real-part of the spectrum is gapped while the imaginary part, controlling the long-time behavior of the non-Hermitian dynamics, is gapless around $k=k_*(h)=\arccos(h)$ where it vanishes linearly $\Gamma_k\sim \alpha_*\vert k-k_*\vert$. In this regime from~\eqref{eqn:Save_dyn} we obtain a logarithmic growth of the entanglement either in time , $\overline{S}_t\sim c_{\rm eff} \log t$, or as a function of the size of the cut $\ell$, $\overline{S}_\infty\sim c_{\rm eff}\log \ell$
with an effective central charge~\cite{alberton2020entanglement,turkeshi2021measurementinduced_1} that we can obtain analytically from ~\eqref{eqn:Save_dyn}. Using as velocity $v_k$ the group velocity of quasi-particles in the initial state we obtain $c_{\rm eff}\sim\sqrt{\gamma_c-\gamma}$, which vanishes as a square root at $\gamma=\gamma_c(h)$~\cite{SM}.  Right at the critical point, $\gamma=\gamma_c(h)$ we have that $\alpha_*\rightarrow \infty$ and the decay is non-analytic $\Gamma_k\sim \vert k-k_*\vert^{1/2}$ which gives a convergent contribution to the entanglement entropy and only leaves a smooth exponential component. On the other hand, for $\gamma>\gamma_c(h)$ a gap opens up in the imaginary part of the quasiparticle spectrum, while the real part is gapless. In this regime, \eqref{eqn:Save_dyn}  leads to an exponential dynamics of the entanglement and to an area law-phase, independent from the size of the cut. 
 As the transverse field is increased above $ h_c =1$ the spectrum changes structure. Specifically, the imaginary part remains always gapped for any $\gamma$ and we, therefore, we do not expect any entanglement transition, a result that we can qualitatively understand by noting that for large $h$ the system remains always very close to an exact eigenstate of the measurement operator and therefore is not able to generate entanglement.

We now present the numerical results obtained by evolving the system under the stochastic dynamics in \eqref{eq:qjump} and compare them with our predictions. In Fig~\ref{fig:dynamics} we plot the average entanglement as a function of time for different values of the transverse field $h$, the measuring strength $\gamma$, and system size $L$~\cite{SM}. We see that for small $h$ and $\gamma$ (top left panel) the entanglement entropy shows the expected logarithmic growth in time, which eventually saturates to a stationary value on time scales that depend on system size. Upon increasing $\gamma$ at a fixed value of $h=0.4$ (top right panel) we see a clear transition in the dynamics of the entanglement which for large $\gamma$ shows an exponentially fast approach to the stationary value and a weak dependence on system size as expected for a Zeno phase. A similar transition in the entanglement dynamics is found for fixed $\gamma=1.5$ upon increasing $h$ (bottom right panel). Finally, for large values of the transverse field, $h=1.5$, and $\gamma=1$ the entanglement entropy reaches a stationary value exponentially fast and no signature of the logarithmic growth phase or transition is found, again in agreement with our phenomenological theory.

In Fig.~\ref{fig:steadystate} we plot (right-panels) the dependence of the stationary entanglement from the system size $L$ for a few representative values of $h,\gamma$, which confirms for $h<h_c=1$ the existence of a logarithmic phase for small $\gamma$, where $\overline{S}_\infty\sim c_{\rm eff}\log L$, and an area law-phase for large $\gamma$, separated by an entanglement transition~\footnote{The central charge $c_\mathrm{eff}$ is obtained through the fit $\overline{S}_\infty=a \log L + b$ neglecting small system sizes and identifying $c_{\rm eff} \equiv a$. }, while for $h>h_c=1$ (bottom panel) only an area law scaling is found, as predicted by \eqref{eqn:Save_dyn}. We note that for certain values of $\gamma$ the stationary entanglement entropy features a crossover with system size, from an area law at small $L$ to a logarithmic scaling at large $L$, missed by our theory which only predicts the large $L$ behavior. A similar crossover was found for the Ising chain in the no-click limit~\cite{turkeshi2021measurementinduced_1} suggesting that the origin of this effect, very different from what usually observed in ground-state critical systems, might be related to the entanglement dynamics of non-Hermitian systems, which is largely unexplored~\cite{bacsi2021dynamics}. Finally, in the left panel we draw a numerical phase diagram in the $(\gamma,h)$ plane by plotting the numerical value of the effective central charge (in the inset we present three representative values of $h$), and compare it with the predicted critical coupling $\gamma_c(h)$ (dashed line) showing a remarkable qualitative agreement. 

Overall these results show that our phenomenological picture captures the essence of the measurement-induced transition in the Quantum Ising model.  We further note that our results for the QJ protocol are qualitatively very similar to those obtained for the same model evolving under continuous monitoring or in the no-click limit~\cite{turkeshi2021measurementinduced_1}, suggesting that our phenomenology might apply more broadly~\footnote{In this context it is worth mentioning that for the free fermionic models considered in Refs.~\cite{alberton2020entanglement}, the presence of a continuous symmetry associated with particle number conservation makes the effective non-Hermitian Hamiltonian somewhat trivial and the no-click limit irrelevant.}.

\emph{Conclusions --- } In this paper we have introduced a phenomenological theory for entanglement transitions 
in quantum many-body systems with well-defined quasi-particles evolving under the effect of continuous monitoring.  The idea is that entanglement is carried by non-Hermitian quasiparticles which propagate ballistically and are reset stochastically with a rate given by  their inverse lifetime. We have written down a renewal equation for the entanglement statistics from which one can obtain the average entanglement in terms of only two parameters, the quasiparticle velocity and their lifetime. We have applied our theory to the Quantum Ising chain measured through quantum jumps. We have shown that our phenomenological theory captures all the key features of this problem, including the logarithmic growth phase, an effective central charge vanishing continuously at the entanglement transition into an area law, and the overall phase diagram. Natural extensions of this work could include monitored free fermionic systems in $d\geq 2$, where one could expect different entanglement scaling to emerge in presence of a gapless spectrum of decay rates, or even interacting integrable quantum many-body systems for which the structure of the effective non-Hermitian Hamiltonian preserve its integrability~\cite{ashida2020nonhermitian}.

\begin{acknowledgements}
This work was supported by the ANR grant ``NonEQuMat''(ANR-19-CE47-0001). We acknowledge computational resources on the Coll\'ege de France IPH cluster. 
\end{acknowledgements}


\newpage

\widetext
\begin{center}
\large{\bf Erratum: `Entanglement Transitions from Stochastic Resetting of Non-Hermitian Quasiparticles' \\}
\end{center}

\author{Xhek Turkeshi}
\affiliation{JEIP, USR 3573 CNRS, Coll\`{e}ge de France, PSL Research University, 11 Place Marcelin Berthelot, 75321 Paris Cedex 05, France}
\author{Marcello Dalmonte}
\affiliation{SISSA, via Bonomea 265, 34136 Trieste, Italy}
\affiliation{ICTP, strada Costiera 11, 34151 Trieste, Italy}
\author{Rosario Fazio}
\affiliation{ICTP, strada Costiera 11, 34151 Trieste, Italy}
\affiliation{Dipartimento di Fisica, Universit\'a di Napoli "Federico II", Monte S. Angelo, I-80126 Napoli, Italy}
\author{Marco Schir\'o}
\affiliation{JEIP, USR 3573 CNRS, Coll\`{e}ge de France, PSL Research University, 11 Place Marcelin Berthelot, 75321 Paris Cedex 05, France}
\maketitle

Our manuscript introduced a phenomenological quasiparticle picture describing monitored many-body systems. 
A central point of our work is that the system's non-Hermitian Hamiltonian (nHH) quasiparticles reveal insights into the measurement-induced phases. 
In particular, the quasiparticle picture explains the emergence of a logarithmic phase in non-interacting monitored fermions when the nHH gap is closed and an area-law phase when the nHH gap is open. (A fact numerically observed in a variety of works (see, e.g., Ref.~\cite{prb1})).

\begin{figure}[h!]
    \centering
    \includegraphics[width=0.7\columnwidth]{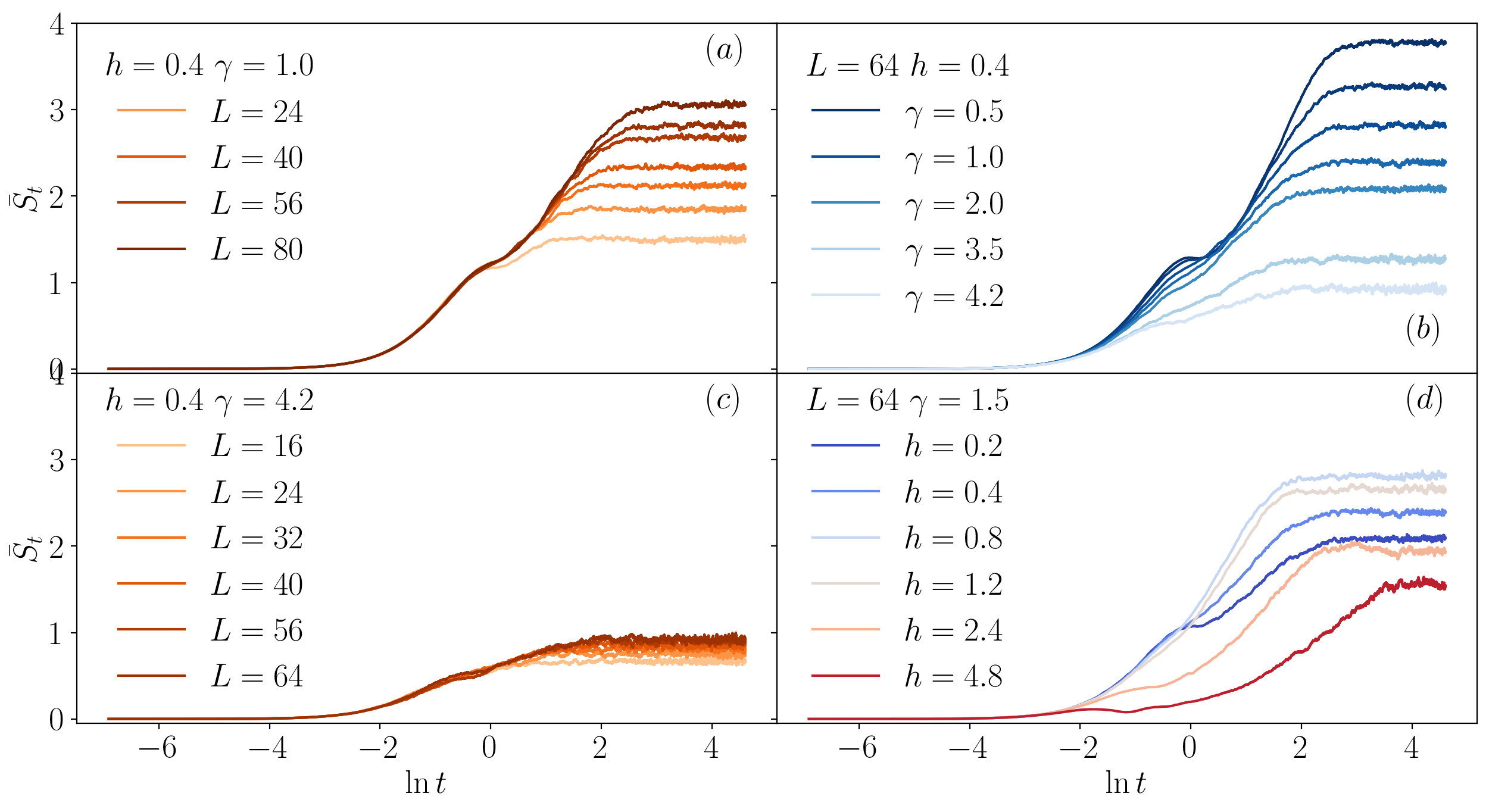}
    \caption{Evolution of the average entanglement entropy for the monitored transverse field Ising model. Logarithmic growth of entanglement in the gapless phase $\gamma<\gamma_c(h) \equiv 4 \sqrt{1-h^2}$ (a), which turns into a system size-independent approach to a constant stationary value upon increasing $\gamma$ above $\gamma_c(h)$ (b). The saturation dynamic for $h>h_c(\gamma)$ shows a weak system size dependence from the system size (c), while upon increasing $h$, a logarithmic growth phase emerges (d). }
    \label{fig:my_label}
\end{figure}

To qualitatively support our claims, we have introduced an archetypal model: the transverse field Ising chain under quantum jumps. Here, the correlation matrix fully captures the dynamics by the system's Gaussianity.

We became aware of an error in our numerical implementation, which we used to extract the data in our manuscript. 
Specifically, the results presented in Fig. 2 and 3 of our original work are incorrect and shall be replaced by the following analysis. Within the computational limitations, we confirm the same qualitative picture of the existence of at least two phases in the phase diagram reported in Fig. 3. However, in the large magnetic field region ($h>1$), we anticipate that the results of the correct simulations are no longer easy to interpret due to considerable finite-size effects.

\begin{figure}[t!]
    \centering
    \includegraphics[width=\columnwidth]{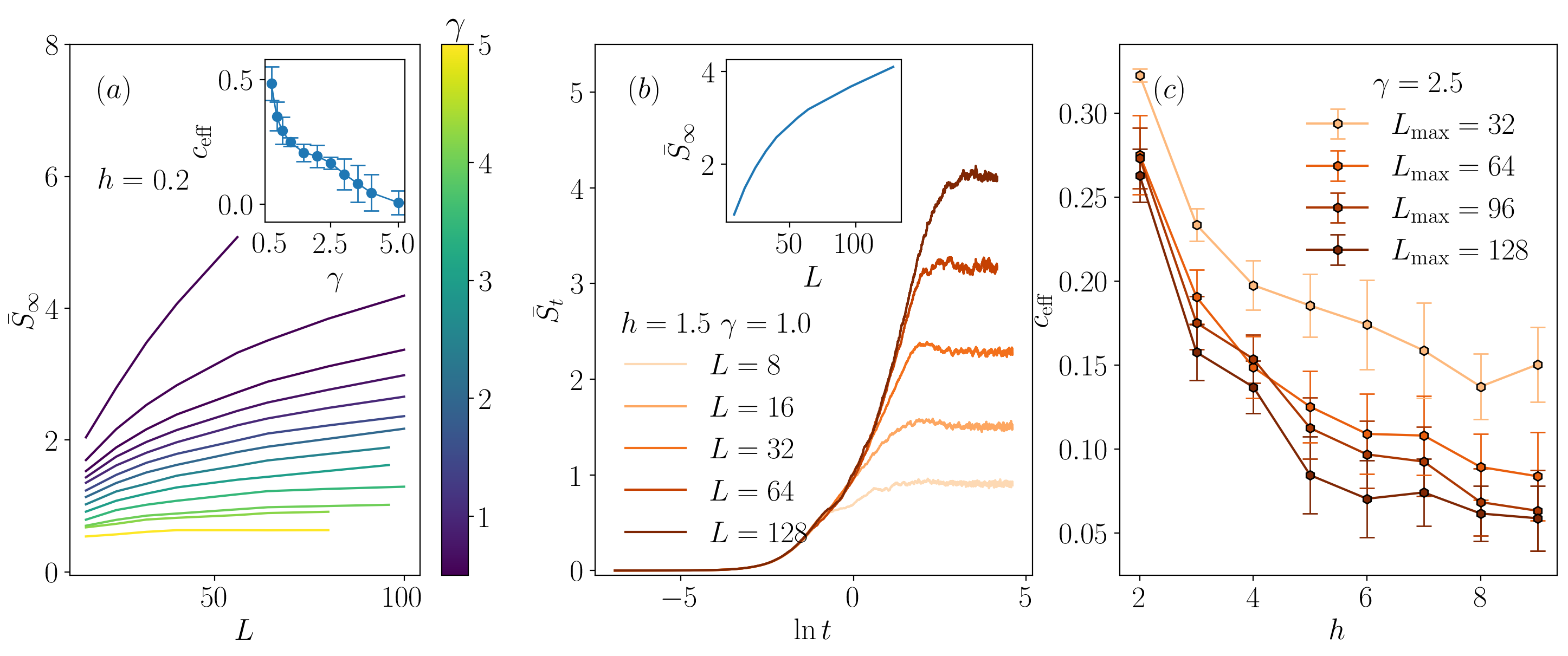}
    \caption{(a) System size scaling and entanglement transition for the stationary state entanglement entropy for $\gamma\in \{0.1,0.3,0.5,0.7,1.0,1.5,2.0,2.5,3,3.5,4,4.2,5\}$. For $\gamma<\gamma_c$, the entanglement entropy exhibits a logarithmic system size scaling, decreasing coefficient as $\gamma$ increases. For $\gamma>\gamma_c$, the system's entanglement saturates to a constant value. Inset: effective central charge varying $\gamma = 0.5\div 5$ using $L_\mathrm{max}\le 100$.  (b) For low measurement rate $\gamma$, the finite size effects are relevant and result in a logarithmic scaling of entanglement even at large $h$. A systematic analysis of these finite-size effects requires novel technological approaches, and it is left for future work. Inset: scaling of the saturation value for various system sizes $L\le 128$. (c) Increasing the system sizes considered, the value of the effective central charge at large $h$ decreases toward smaller values.  }
    \label{fig:my2}
\end{figure}

The updated numerical analysis is given in Fig.~\ref{fig:my_label}-\ref{fig:my2}. First, we notice the existence of a phase with logarithmic growth of entanglement entropy for $\gamma<\gamma_c=4\sqrt{1-h^2}$, as predicted by the quasiparticle picture, see Fig. 1(a). At a fixed $h$, the entanglement growth decreases for larger values of $\gamma$ (cf. Fig. 1(b)). When the measurement rate exceeds the critical threshold $\gamma_c$, the system's entanglement entropy quickly saturates to a weakly system size-dependent value, see Fig. 1(c). 
Similarly to the data in the original manuscript, the finite size effect presents larger entropy close to $h=1$ (where the unitary evolution is the most scrambling ~\cite{calabrese2005evolution}); see Fig. 1(d). However, these effects are reduced by considering larger system sizes. 

Furthermore, the saturated entanglement entropy displays a logarithmic growth for $\gamma<\gamma_c(h)$ and saturation to a constant value for $\gamma>\gamma_c(h)$, see Fig.~\ref{fig:my2}(a). As detailed in the manuscript, this effect reflects the gap opening in the imaginary part of the nHH. In the inset, we include the scaling of the effective central charge, fitted using $\bar{S}_\infty = c_\mathrm{eff} \ln L + b$. At lower values of $\gamma$, $c_\mathrm{eff}$ grows and, for sufficiently low measurement rate $\gamma$, the growth of $\bar{S}_\mathrm{\infty}$ is linear in system size, a symptom of a spurious volume law behavior due to finite sizes~\cite{alberton2020entanglement,fidkowski2021howdynamicalquantum}. 

In summary, the main difference compared to the results in the original manuscript is the presence of more significant finite-size effects affecting the entanglement entropy dynamics. In particular, for $\gamma$ sufficiently small and $h$ larger than one, the accessible system sizes display a logarithmic growth of entanglement (cf. Fig.~\ref{fig:my2}(b); In the inset, we show that also the saturation value follows a logarithmic scaling in system size.)
We attribute this phenomenon to finite-size effects, which can be expected to be important in this part of the phase diagram where the gap in the nHH, and thus the resetting rate of the quasiparticles, is given by $\gamma\ll h$ and thus is the smallest scale in the problem. Indicatively, in Fig.~\ref{fig:my2}(c), we demonstrate that considering larger system sizes for the fit of the central charge, the effective central charge decreases to smaller values. This decrease is in qualitative agreement with the quasiparticle picture, which would expect $c_\mathrm{eff}=0$ in the thermodynamic limit for $h>1$. 

Nevertheless, with the present numerical results, we cannot draw a precise phase diagram as that in Fig.~3 of our previous manuscript. In particular, the available numerical methods cannot exclude novel physical phenomenology beyond the quasiparticle picture for large magnetic field $h$ and low measurement rate. A definitive analysis of this issue requires novel simulation methods, which go beyond our manuscript's scope and are left for future investigation.

In conclusion, the new analysis confirms the qualitative description given by the quasiparticle picture for monitored fermionic systems in a wide range of parameters, provided finite-size effects are considered.

We are grateful to A. Paviglianiti and A. Silva for pointing out a problem in our original numerical implementation.

%

\begin{center}
\large{\bf Supplemental Material to `Entanglement Transitions from Stochastic Resetting of Non-Hermitian Quasiparticles' \\}
\end{center}

\author{Xhek Turkeshi}
\affiliation{JEIP, USR 3573 CNRS, Coll\`{e}ge de France, PSL Research University, 11 Place Marcelin Berthelot, 75321 Paris Cedex 05, France}
\author{Marcello Dalmonte}
\affiliation{SISSA, via Bonomea 265, 34136 Trieste, Italy}
\affiliation{ICTP, strada Costiera 11, 34151 Trieste, Italy}
\author{Rosario Fazio}
\affiliation{ICTP, strada Costiera 11, 34151 Trieste, Italy}
\affiliation{Dipartimento di Fisica, Universit\'a di Napoli "Federico II", Monte S. Angelo, I-80126 Napoli, Italy}
\author{Marco Schir\'o}
\affiliation{JEIP, USR 3573 CNRS, Coll\`{e}ge de France, PSL Research University, 11 Place Marcelin Berthelot, 75321 Paris Cedex 05, France}

\maketitle

In this Supplemental Material, we provide details on (i) the derivation of the renewal equation for the entanglement probabilty, (ii) an alternative formulation in terms of last-resetting renewal equation, (iii) the explicit solution of the renewal equation, (iv) the spectrum of the non-Hermitian Ising chain and the scaling of the average entanglement (v) the quantum jumps protocol for the quantum Ising chain and its numerical implementation.

\section{Renewal Equation for Entanglement Probability}

Here we show how to derive the renewal equation given in Eq. (3) of the main text from a  resummation of resetting events~\cite{evans2020stochastic}. For the sake of generality we consider resetting events to be distributed according to a waiting-time probability $\varphi_k(t)$ which gives the probability of having a resetting event in the interval $t,t+dt$.
By construction, this has to be normalized
\begin{equation}
\int_0^{\infty}dt\varphi_k(t)=1
\end{equation}
and for a Poisson process, considered in the main text, this reduces to $\varphi_k(t)=\Gamma_k\,e^{-\Gamma_k t}$. As a consequence the probability of no resetting up to time $t$ is 
\begin{equation}
	\Psi_k(t)=1-\int_0^t dt'\psi_k(t')\equiv e^{-\Gamma_k t}
\end{equation}
where $ \int_0^t dt'\psi_k(t')=e^{-\Gamma_k t}-1$ is the probability of having at least one resetting in the time interval $(0,t)$. In terms of these quantities the renewal equation given in the main text, also called \emph{first-resetting renewal equation}, reads
\begin{align}\label{eqn:first-time-renewal_SM}
P_k(s,t)=\Psi_k(t)p_k^0(s,t)+\int_0^t d\tau \varphi_k(\tau)P_k(s,t-\tau)
\end{align}
where we recall that $p_k^0(s,t)$, given in the main text, is the probability of an entanglement contribution $s$ in absence of any resetting.

In order to derive Eq.~(\ref{eqn:first-time-renewal_SM}) we write the probability $P_k(s,t)$ as a series expansion over an arbitrary number $n$ of resettings events, i.e.
\begin{equation}\label{eqn:series_Pkst}
	P_k(s,t)=\sum_{n=0}^{\infty}P^{(n)}_k(s,t)\,,
\end{equation}
where $P^{(n)}_k(s,t)$ is the probability of having an entanglement contribution $s$ at time $t$ in presence of $n$ resettings. We start to evaluate explicitly the first terms of this series. For $n=0$, corresponding to no resettings up to time $t$, we have by definition
\begin{equation}
	P_k^{(n=0)}(s,t)=\Psi_k(t)p^0_k(s,t)\;
\end{equation}
For $n=1$ we have one resetting at time $t_1\in (0,t)$, followed by no resetting in a time interval of duration $t-t_1$ with free propagation starting from $s^*=0$ for a time $t-t_1$. This gives a contribution
\begin{equation}
P_k^{(n=1)}(s,t)=\int_0^t dt_1\varphi_k(t_1)\Psi_k(t-t_1)p^0_k(s,t-t_1)\;
\end{equation}
Similarly, for $n=2$ we can have one resetting at time $t_1\in (0,t)$ and a second after a time $t_2\in (0,t-t_1)$, followed by no resetting in a time interval of duration $t-t_1-t_2$ with free propagation starting from $s^*=0$ for a time $t-t_1-t_2$, with contribution
\begin{equation}
P_k^{(n=2)}(s,t)=\int_0^t dt_1\varphi_k(t_1)\int_0^{t-t_1} dt_2\varphi_k(t_2)
\Psi_l(t-t_1-t_2)p_k^0(s,t-t_1-t_2)\;
\end{equation}
This procedure can be straightforwardly iterated to any order $n$ to obtain the entanglement probability in presence of $n$ resettings.
It is now easy to see that the renewal equation given in Eq.~(\ref{eqn:first-time-renewal_SM}) generates the very same expansion. 
In fact we take  Eq.~(\ref{eqn:first-time-renewal_SM}) for  $P_k(s,t)$ and we use it to express the probability at time $t-\tau$, i.e. $P_k(s,t-\tau)$ entering the right hand side, we obtain
\begin{equation}
P_k(s,t-\tau)=\Psi_k(t-\tau)p_k^0(s,t-\tau)+\int_0^{t-\tau} d\tau_1 \varphi_k(\tau_1)P_k(s,t-\tau-\tau_1)
\end{equation}
Plugging this result in Eq.~(\ref{eqn:first-time-renewal_SM}) we obtain
\begin{equation}
P_k(s,t)=\Psi_k(t)p_k^0(s,t)+\int_0^t d\tau \varphi_k(\tau)\Psi_k(t-\tau)p_k^0(s,t-\tau)+
\int_0^t d\tau\,\varphi_k(\tau)\int_0^{t-\tau} d\tau_1 \varphi_k(\tau_1)P_k(s,t-\tau-\tau_1).
\end{equation}
If we stop at this order and identify the distribution $P_k(s,t-\tau-\tau_1)$ appearing on the right hand side as the bare one, i.e.
\begin{equation}
P_k(s,t-\tau-\tau_1)=\Psi_k(t-\tau-\tau_1)p_k^0(s,t-\tau-\tau_1)
\end{equation}
we obtain the resummation of resetting events up to $n=2$ that we discussed above. However we can continue this process, using Eq.~(\ref{eqn:first-time-renewal_SM}) to generate a full series of nested integrals which reproduces the series expansion in resetting events to all orders. We conclude therefore that the renewal equation provides a formal resummation of resetting events of arbitrary number.

\section{Alternative Formulation: Last-Resetting Renewal Equation}

Here we show that the series expansion in Eq.~(\ref{eqn:series_Pkst}), or equivalently the first-resetting renewal equation in Eq.~(\ref{eqn:first-time-renewal_SM}), can be in fact resummed analytically in a closed form known as \emph{last-resetting renewal equation}~\cite{evans2020stochastic}, which reads
\begin{align}\label{eqn:last_renewal_general_SM}
P_k(s,t)=e^{-\Gamma_kt}\,p_k^0(s,t)+\Gamma_k\int_0^t d\tau e^{-\Gamma_k\tau}p_k^0(s,\tau)
\end{align}
To see this it is convenient to express the series in Eq.~(\ref{eqn:series_Pkst}) in Laplace domain. Specifically we can introduce the Laplace transform
\begin{align}
\tilde{P}_k(s,z)=\int_0^{\infty}dt e^{-zt}P_k(s,t)=\sum_{n=0}^{\infty}\tilde{P}^{(n)}_k(s,z)\,,
\end{align}
and evaluate each term in the expansion. For the $n=1$ term in Laplace space we get
\begin{align}
\int_0^{\infty} dt e^{-zt}\int_0^t dt_1\varphi_k(t_1)\Psi_k(t-t_1)p^0_k(s,t-t_1)&=
\int_0^{\infty} dt_1\,\int_{t_1}^{\infty}dt e^{-zt}\varphi_k(t_1)\Psi_k(t-t_1)p^0_k(s,t-t_1)\nonumber \\
&=\int_0^{\infty} dt_1\,\int_{0}^{\infty}d\tau e^{-z(t_1+\tau)}\varphi_k(t_1)\Psi_k(\tau)p_k^0(s,\tau)=
\tilde{\varphi}_k(z)\tilde{F}_k(s,z)
\end{align}
where $\tilde{\varphi}_k(z)$ is the Laplace transform of $\varphi_k(t)$ and we have defined $\tilde{F}_k(s,z)$
\begin{align}
\tilde{F}_k(s,z)=\int_0^{\infty}d\tau\,e^{-z\tau}\Psi_k(\tau)p_k^0(s,\tau)
\end{align}
With similar manipulations we can get for $n=2$
\begin{align}
P_k^{(n=2)}(s,z)=\int_0^{\infty} dt e^{-zt}\int_0^t dt_1\varphi_k(t_1)\int_0^{t-t_1} dt_2\varphi_k(t_2)
\Psi_l(t-t_1-t_2)p_k^0(s,t-t_1-t_2)=\tilde{\varphi}^2_k(z)\tilde{F}_k(s,z)
\end{align}
If we iterate this procedure and resum the geometric series we obtain
\begin{align}\label{eqn:Pksz}
\tilde{P}_k(s,z)=\tilde{F}_k(s,z)+\tilde{F}_k(s,z)\sum_{n=1}^{\infty}\tilde{\varphi}_k(z)^n=
\tilde{F}_k(s,z)+\tilde{F}_k(s,z)\frac{\tilde{\varphi}_k(z)}{1-\tilde{\varphi}_k(z)}=
\frac{\tilde{F}_k(s,z)}{1-\tilde{\varphi}_k(z)}
\end{align}
From the solution in Laplace space we can go back to the real-time. It is easy to see that the function
\begin{align}\label{eqn:last_renewal_general}
P_k(s,t)=\Psi_k(t)p_k^0(s,t)+\int_0^t d\tau \Upsilon_k(\tau)\Psi_k(t-\tau)p_k^0(s,t-\tau)
\end{align}
gives back Eq.~(\ref{eqn:Pksz}) under Laplace transform, provided the Laplace transform of the function $\Upsilon_k(t)$ is such that
\begin{align}\label{eqn:Upsilonk}
\tilde{\Upsilon}_k(z)=\int_0^{\infty} dt\,e^{-zt}\Upsilon_k(t)\equiv \frac{\tilde{\varphi}_k(z)}{1-\tilde{\varphi}_k(z)}
\end{align}
In fact we have
\begin{align}
\int_0^{\infty} dt e^{-zt}P_k(s,t)&=\tilde{F}_k(s,z)+\int_0^{\infty}  dt e^{-zt}\int_0^t d\tau \Upsilon_k(\tau)\Psi_k(t-\tau)p_k^0(s,t-\tau)\nonumber\\
&=\tilde{F}_k(s,z)+\int_0^{\infty} d\tau\int_t^{\infty}  dt e^{-zt} \Upsilon_k(\tau)\Psi_k(t-\tau)p_k^0(s,t-\tau)\nonumber\\
&=\tilde{F}_k(s,z)+\int_0^{\infty} d\tau\int_0^{\infty}  dt' e^{-z(t'+\tau)} \Upsilon_k(\tau)\Psi_k(t')p_k^0(s,t')\nonumber\\
&=\tilde{F}_k(s,z)+\tilde{\Upsilon}_k(z)\tilde{F}_k(s,z)
\end{align}
which coincides with Eq.~(\ref{eqn:Pksz}) provided that Eq.~(\ref{eqn:Upsilonk}) holds. For Poissonian resetting we can further simplify Eq.~(\ref{eqn:last_renewal_general}) since we have

\begin{align}
&\varphi_k(t)=\Gamma_k\,e^{-\Gamma_kt}\\
&\Psi_k(t)=e^{-\Gamma_kt}
\end{align}
and in Laplace domain
\begin{align}
&\tilde{\varphi}_k(z)=\frac{\Gamma_k}{\Gamma_k+z}\\
&\tilde{\Psi}_k(z)=\frac{1}{\Gamma_k+z}
\end{align}
from which we obtain $\tilde{\Upsilon}_k(z)=\Gamma_k/z$ which in the time domain gives $\Upsilon_k(t)=\Gamma_k\theta(t)$ and allow us to rewrite Eq.~(\ref{eqn:last_renewal_general}) as 
\begin{align}
P_k(s,t)=e^{-\Gamma_kt}\,p_k^0(s,t)+\Gamma_k\int_0^t d\tau e^{-\Gamma_k\tau}p_k^0(s,\tau)
\end{align}
which coincides with Eq.~(\ref{eqn:last_renewal_general_SM}) which therefore provides a closed form solution to the first-time renewal equation.

\section{Solution of Renewal Equation and Average Entanglement}

Once the probability of the entanglement contribution $P_k(s,t)$ is known from Eq.~(\ref{eqn:last_renewal_general_SM}) we can obtain all the moments of the entanglement contribution of each momentum $k$
\begin{equation}
	\overline{s^n_k}(t)\equiv\int ds P_k(s,t) s^n
\end{equation}
as well as the associated moments of the entanglement entropy
\begin{equation}
\overline{S^n}(t)\equiv\int \frac{dk}{2\pi} \overline{s^n_k}(t)
\end{equation}
In order to obtain an explicit expression for $P_k(s,t)$ it is useful to introduce its characteristic function $F_k(q,t)$  defined as 
\begin{align}
P_k(s,t)=\int dq e^{iqs}F_k(q,t)\\
F_k(q,t)=\int dse^{-iqs} P_k(s,t)
\end{align}
The logarithm of the characteristic function $\Phi_k(q,t)=\mbox{log}F_k(q,t)$ generates all the connected moments of the entanglement entropy contribution of each quasiparticle. 
We can define therefore the characteristic function of the full entanglement distribution 
\begin{align}\label{eqn:Phiq}
\Phi(q,t)=\int \frac{dk}{2\pi}\Phi_k(q,t)
\end{align}
which by construction generates the connected moments of the entanglement entropy, for example for the average we have
\begin{align}\label{eqn:S_ave}
\partial \Phi/\partial q\vert_{q=0} &= -i \overline{S(t)}
\end{align}
We can now compute explicitly the characteristic function $F_k(q,t)$. Plugging the renewal equation into the definition we have
\begin{equation}
F_k(q,t)=e^{-\Gamma_k t}e^{-iqS_k^*(t)}+\Gamma_k\int_0^t d\tau e^{-\Gamma_k\tau}e^{-iqS_k^*(\tau)}.
\end{equation}
In order to evaluate the integral on the right hand side we have to distinguish whether $t$ is smaller or greater than the time scale $t^*=l/2\vert v_k\vert$. Specifically for $t<t^*$ we have that also $\tau<t^*$ and we obtain
\begin{equation}
F_k(q,t)=e^{-\Gamma_k t}e^{-iqs_02\vert v_k\vert t}+\Gamma_k\int_0^t d\tau e^{-\Gamma_k\tau}e^{-iqs_02\vert v_k\vert \tau }
\end{equation}
while for  $t>t^*$ we have two different contributions to the integral
\begin{equation}
F_k(q,t)=e^{-\Gamma_k t}e^{-iqs_0 l}+\Gamma_k\int_0^{t^*} d\tau e^{-\Gamma_k\tau}e^{-iqs_02\vert v_k\vert \tau }+
\Gamma_k\int_{t^*}^t d\tau e^{-\Gamma_k\tau}e^{-iqs_0l }.
\end{equation}
Performing the integrals and after simple manipulation we end up with the following result for the characteristic function
\begin{align}
F_k(q,t)=\frac{\Gamma_k+iqB_k(t) e^{-iq \tilde{B}_k \mbox{min}(t,t^*)} }{\Gamma_k+iq\tilde{B}_k}
\end{align}
where we have defined 
\begin{align}\label{eqn:btilde}
\tilde{B}_k&=2s_0\vert v_k\vert\\
\label{eqn:bt}
B_k(t)&=\tilde{B}_k\,e^{-\Gamma_k\mbox{min}(t,t^*)} .
\end{align}

We can now evaluate the average entanglement entropy as a function of time. To this extent we use Eqns.~(\ref{eqn:Phiq}), (\ref{eqn:S_ave}), and take the derivative with respect to $q$ to get
\begin{align}
 \overline{S} = i \partial \Phi/\partial q\vert_{q=0}=\int \frac{dk}{2\pi}\frac{1}{F_k(q=0)}\frac{i\partial F_k}{\partial q}\vert_{q=0}
\end{align}
A simple calculation for the first derivative evaluated at $q=0$ gives the result 
\begin{equation}
	\frac{\partial F_k}{\partial q}\vert_{q=0}=\frac{i\left(B_k(t)-\tilde{B}_k\right)}{\Gamma_k}
\end{equation}
from which we immediately recover Eq. (6) of the main text, i.e.
\begin{equation}\label{eqn:Save_dyn_SM}
\overline{S}_t(\ell)=
\int \frac{dk}{2\pi}
\frac{2\vert v_k\vert}{\Gamma_k}
\left[1-e^{-\Gamma_k \mbox{min}(t,t^*)}
\right]\;.
\end{equation}

\section{Spectrum of Non-Hermitian Ising Model and Scaling of  Entanglement Entropy}

In the main text we consider the following non-Hermitian Ising chain
\begin{equation}
H_{\rm eff}=-J\sum_i \sigma^x_i\sigma^x_{i+1}-h\sum_i\sigma^z_i-i\frac{\gamma}{4}\sum_i \sigma^z_i
\end{equation}
which is obtained in the no-click limit of the QJ protocol discussed in the main text, using the fact that $L_i=(1+\sigma^z_i)/2\equiv L_i^2$ and dropping the constant factor. Using a Jordan-Wigner transformation of spins into fermions this model can be diagonalised exactly in terms of new fermionic degrees of freedoms $\hat{\bar{\gamma}}_k,\hat{\gamma}_k$~\cite{hickey2013timeintegrated,lee2014heralded,biella2021manybody} and brought into a form 
\begin{equation}
H_{\rm eff}=\sum_{k>0} \Lambda_k\left(\hat{\bar{\gamma}}_k\hat{\gamma}_k+\hat{\bar{\gamma}}_{-k}
\hat{\gamma}_{-k}-1\right)
\end{equation}
where the complex quasiparticle spectrum $\Lambda_k$ reads
\begin{equation}\label{eqn:Lambdak_SM}
	\Lambda_k= \sqrt{\varepsilon_k^2-\gamma^2/4+2i\gamma\left(h-\cos k\right)}
\end{equation}
where $\varepsilon_k=\sqrt{4\left(h^2+1-2h\cos k\right)}$ is the dispersion of the Hermitian Ising chain and we fixed $J=1$. We define the real and imaginary part of the complex eigenvalues $\Lambda_k$ in such a way that the imaginary part is always positive (decays modes).
\begin{equation}
	\Lambda_k=E_k+i\Gamma_k.
\end{equation}
As discussed in the main text, the scaling of the average entanglement entropy predicted from the stochastic resetting theory is directly controlled by the spectral property of the decay modes, encoded in the quasiparticle lifetime $\Gamma_k$. By inspecting Eq.~(\ref{eqn:Lambdak_SM}) we immediately see that two cases have to be distinguished, depending on the value of $h$. 
For $h<h_c=1$ the imaginary part of the argument of the square root in Eq.~(\ref{eqn:Lambdak_SM}) vanishes at a pair of $k$ points given by $k=k_*(h)=\pm \arccos(h)$. Then we see that at these values of $k$ the spectrum reads $\Lambda_{k_*}=\sqrt{4(1-h^2)-\gamma^2/4} $ and it is therefore either purely real, when
\begin{equation}\label{eqn:sub}
4\left(1-h^2\right)>\gamma^2/4
\end{equation}
or purely imaginary otherwise. The condition above defines in fact a spectral phase transition in the non-Hermitian Hamiltonian for measurement strength $\gamma$ equal to the critical value $\gamma_c(h)=4\sqrt{1-h^2}$. 

For $\gamma<\gamma_c(h)$ we have
$$
E_{k_*}=\sqrt{\gamma_c^2/4-\gamma^2/4}\;\quad\;\Gamma_{k_*}=0
$$ 
while for  $\gamma>\gamma_c(h)$ we have
$$
E_{k_*}=0\;\quad\;\Gamma_{k_*}=\sqrt{\gamma^2/4-\gamma_c^2/4}
$$ 
Near  $k_*$ we can linearize the spectrum and find that  for $\gamma<\gamma_c(h)$ the decay rate $\Gamma_k$ behaves as
\begin{equation}\label{eqn:gamma_k_linear}
\Gamma_k\simeq \alpha_*\vert k-k_*\vert
\end{equation}
where $\alpha_*$ can be obtained from the derivative of $\Lambda_k$ and reads
$$
 \alpha_*=\mbox{Im}\frac{d\Lambda}{dk}\vert_{k=k_*}=\mbox{Im}\frac{1}{2\Lambda_{k^*}}\left(2\varepsilon_{k_*}\frac{d\varepsilon}{dk}\vert_{k=k_*}+2i\gamma\sin k_*\right)=2\gamma\sqrt{1-h^2}/\sqrt{\gamma_c^2-\gamma^2}
$$
As the critical point is approached, $\gamma\rightarrow\gamma_c$, we have that $\alpha_*\rightarrow\infty$ and the decay rate is non analytic, $\Gamma_k\sim \vert k-k_*\vert^{1/2}$. On the other hand for $h>h_c=1$ the imaginary part of the argument of the square root in Eq.~(\ref{eqn:Lambdak_SM}) is always different from zero and as such there is always a finite decay rate in the system.

Using the results above and Eq.(\ref{eqn:Save_dyn_SM}) we can obtain the predictions for the average entanglement entropy discussed in the main text. In particular for $\gamma<\gamma_c$, when Eq.~(\ref{eqn:gamma_k_linear}) holds, we have
$$
\overline{S}_t(\ell)=
\int \frac{dk}{2\pi}
\frac{2\vert v_k\vert}{\Gamma_k}
\left[1-e^{-\Gamma_k \mbox{min}(t,t^*)}
\right]
\simeq \frac{\vert v_{k_*}\vert}{\pi \alpha_*}\int \frac{dk}{k}\left(1-\exp(-\alpha_*\,k\,\mbox{min}(t,t^*)\right)
$$
which we see is infrared divergent in the large $\ell,t$ limit and leads to a logarithmic scaling of the entanglement entropy, $\overline{S}_t(\ell)\sim c_{\rm eff}\log \ell$  with a prefactor
$$
c_{\rm eff}=\frac{1}{\gamma\pi}\sqrt{\gamma_c^2-\gamma^2}
$$
where we have used as quasiparticle velocity $v_k$  the group velocity of quasi-particles in the initial state, $v_k=d\varepsilon_k/dk= 4h_0\sin k/\varepsilon_k(h_0)\sim 2\sin k$ for $h_0\gg 1$.

\section{Details on Numerical Simulations: Quantum Jump Equation and Entanglement Evolution}
In this section we consider the evolution of the TFIM subject to quantum jumps. We first introduce the equation of motion for the quantum state in the Hilbert space, which we then recast in an evolution for the two-point correlation function. Throughout this section we simplify the notation introducing $|\psi_t\rangle\equiv |\psi(\boldsymbol{\mathcal{N}}_t)\rangle$.

\textit{Stochastic Schr\"odinger equation for the quantum state -- } The state dynamics is given by
\begin{align}
	d|\psi_t\rangle = -i H dt |\psi_t\rangle -\frac{\gamma}{2}dt \sum_l (n_l-\langle n_l\rangle_t) |\psi_t\rangle + \sum_l d\mathcal{N}_{l,t}\left(\frac{n_l}{\sqrt{\langle n_l\rangle_t}}-1\right)|\psi_t\rangle,\label{eq:sse}
\end{align}
where $H=-\sum_{i} (J\sigma^x_i \sigma^x_{i+1} + h \sigma_i^z)$ is the Ising Hamiltonian, $\sigma^\alpha_l$ are Pauli matrices acting on site $l$,  $n_l=(1+\sigma_l^z)/2= c^\dagger_l c_l$ is the number operator acting on site $l$, and $\langle O\rangle_t\equiv \langle\psi_t|O|\psi_t\rangle$ is the expectation value at time $t$.
The trajectory nature of Eq.~\eqref{eq:sse} is fully captured in the Poisson process $d\mathcal{N}_{l,t}=0,1$ with $d\mathcal{N}_{l,t} d\mathcal{N}_{l',t} = \delta_{l,l'} d\mathcal{N}_{l,t}$ and disorder average value $\overline{d\mathcal{N}_{l,t}} = \gamma dt \langle n_l\rangle_t$.

Eq.~\eqref{eq:sse} preserves the Gaussianity: given an initial state $|\psi_0\rangle$ in the fermionic operators $c$ and $c^\dagger$, the state state at time $t$ is also Gaussian.
A major consequence is that the state is fully encoded in the two-body correlation functions of the fermionic operators, \emph{i.e.} in the $2L\times 2L$ matrix 
\begin{align}
	\mathbb{G}(t) \equiv \begin{pmatrix}
		C(t)& F(t)\\ F^\dagger(t) & 1-C^T(t)
	\end{pmatrix}\;,\label{eq:gmat}
\end{align}
with the submatrices given by
\begin{align}
	C_{mn}(t)\equiv \langle c_m c^\dagger_n\rangle_t\; ,\qquad F_{mn}(t) \equiv \langle c_m c_n\rangle_t\; .
\end{align}
Furthermore, the entropy can be easily derived from Eq.~\eqref{eq:gmat}. 
Introducing the matrix
\begin{equation}
	\mathbb{W} = \begin{pmatrix}
		\mathbf{1}_{L\times L} & \mathbf{1}_{L\times L}\\
		-i \mathbf{1}_{L\times L} & i \mathbf{1}_{L\times L}
	\end{pmatrix}\;,
\end{equation}
where $\mathbf{1}_{L\times L}$ is the $L\times L$ identity matrix, we define the matrix (\emph{en passant}, we note this is the correlation matrix for Majorana fermions)
\begin{align}
	\mathbb{M}(t) = \mathbb{W}\mathbb{G}(t) \mathbb{W}^\dagger.
\end{align}
The entanglement entropy of a connected partition $A=\{1,2,\dots,\ell\}$ is encoded in the $2\ell\times 2\ell$ matrix $\mathbb{A}_{l,l'}$ read from
\begin{align}
	\mathbb{M}_{l,l'} = \delta_{l,l'} + i \mathbb{A}_{l,l'}\;, & \qquad &\mathbb{M}_{l,l'+L} = i \mathbb{A}_{l,l'+\ell}\;,\\
	\mathbb{M}_{l+L,l'+L} = \delta_{l,l'} + i \mathbb{A}_{l+\ell,l'+\ell}\;, & \qquad & \mathbb{M}_{l+L,l'} = i \mathbb{A}_{l+\ell,l'}.
\end{align}
The matrix $\mathbb{A}$ is real and anti-symmetric and can be reduced to canonical form through Schur's decomposition
\begin{align}
	\mathbb{A} = \mathbb{R} \mathbb{J} \mathbb{R}^T,\qquad \text{with }\quad \mathbb{J} = \bigoplus_{l=1}^\ell \begin{pmatrix}
		0 & \lambda_l\\
		-\lambda_l & 0,
	\end{pmatrix}
\end{align} 
with $\lambda_l\in [-1,1]$. Given $\eta_l(t) = (1+\lambda_l(t))/2$ and $1-\eta_l(t) = (1-\lambda_l(t))/2$, the entanglement entropy is given the Yang-Yang entropy
\begin{align}
	S_A(t) = -\sum_l ( \eta_l \ln \eta_l + (1-\eta_l) \ln(1-\eta_l)).
\end{align}

\textit{Stochastic Schr\"odinger equation for the correlation functions -- }
In this paragraph we derive the equation of motion for $\mathbb{G}(t)$. We preliminary note the stochastic Schr\"odinger equation Eq.~\eqref{eq:sse} can be divided in two parts: a deterministic evolution, generated by the non-Hermitian Hamiltonian 
\begin{equation}
	H_\mathrm{eff} = H - i\frac{\gamma}{2} dt \sum_l n_l\;,
\end{equation}
and by quantum jumps 
\begin{equation}
	|\psi_t\rangle \mapsto |\psi'_t\rangle = \frac{n_l}{\sqrt{\langle n_l\rangle_t}} |\psi_t\rangle.
\end{equation}
Since the measurement nature is strong ($d\mathcal{N}=1\gg dt$ and $n_l$ is a projection), we can separately consider a non-Hermitian evolution for $\mathbb{G}$ interspersed with stochastic measurements (see Ref.~\cite{Daley2014}).

The equation of motion for the matrices $C_{m,n}=\langle {c}_m {c}_n^\dagger\rangle$ and $F_{m,n} = \langle {c}_m {c}_n\dagger\rangle$ are obtained using the Jordan-Wigner transformation and using Wick theorem. They are given by 
 \begin{align}
     dC_{m,n} &= -2i dt \sum_{l} (A_{m,l} C_{l,n}- C_{m,l}A_{l,n} + B_{m,l}(F^\dagger)_{l,n} \nonumber\\
         & + F_{m,l}B_{l,n}) - \gamma dt C_{m,n} + \gamma dt \sum_l (C_{m,l} C_{l,k} + F^*_{n,l} F_{l,m}) + \sum_{l} dN_l \mathbb{J}^C_l [C_{m,n}]\\
     dF_{m,n} &= -2i dt \sum_{l} (A_{m,l} F_{l,n} + B_{m,l}(\delta_{l,n} - C_{n,l}) \nonumber  \\ & 
         + F_{m,l} A_{l,n} - C_{m,l} B_{l,n}) - \gamma dt F_{m,n} - \gamma dt\sum_{l}(F_{l,m} C_{n,k} - F_{k,n} C_{m,k}) + \sum_{l} dN_l \mathbb{J}^F_l [F_{m,n}]\\
     d\mathbb{J}^C_l[C_{m,n}] &= - \frac{C_{m,l} C_{l,n}}{C_{l,l}} + \delta_{m,l}\delta_{l,n} - \frac{F_{l,m}F_{n,l}^*}{C_{l,l}} \qquad 
     d\mathbb{J}^F_l[F_{m,n}] = - \frac{C_{m,l} F_{l,n}}{C_{l,l}} + \frac{C_{n,l}F_{l,m}}{C_{l,l}}.\label{eq:finalgevo}
 \end{align}
 In the above equations, the matrix $A$ and $B$ specify the Hamiltonian and are given by $A_{m,m+1}=A_{m+1,m}=-J/2$, $A_{m,m}=-h$, and $B_{m,m+1}= -B_{m+1,m} = -J/2$. 
 
 \paragraph{Numerical Implementation --} We numerically implement Eq.~\eqref{eq:finalgevo} following the prescription in Ref.~\cite{Daley2014}. 
The transition probability in a small time-step $\delta t$ is given by the decay of the norm
\begin{align}
	\delta p = \delta t \langle \psi_t | i (H_\mathrm{eff} - H_\mathrm{eff}^\dagger) |\psi_t\rangle = \delta t \sum_l \gamma \langle n_l\rangle \equiv  \sum_l \delta p_l.
\end{align}
We can interpret $\delta p_l$ as the probability that the action of $n_l$ will occur during this particular time step. 
We draw a uniform random number $r_1\in [0,1]$, and compare it with $\delta p$. If $r_1>\delta p$, no jump occurs.

Instead, if $r_1<\delta p$ a jump occur. To identify the site where the jump occurs, we draw a random number $r_2$ and we split the interval $[0,1]$ in $L$ sub-intervals $\cup_{m}[\sum_{l=0}^{m-1} \delta \tilde{p}_l,\sum_{l=0}^m \delta \tilde{p}_l]$, where $\delta \tilde{p}_l\equiv \delta p_l/\delta p$, $m=1,\dots L$, and  $\delta \tilde{p}_0\equiv 0$. The qubit undergoing the jump is then identified by the $m$-th sub-interval $r_2 \in [\sum_l^{m-1} \delta \tilde{p}_l,\sum_l^{m} \delta \tilde{p}_l]$.

For the numerical simulations we fix $J=1$ and consider the average over $N_\mathrm{realiz}=300\div 1000$ disorder realizations and $dt\sim (L\gamma)^{-1}$ (with a minimal choice of $dt=0.01$) to obtain accurate entanglement growth. Furthermore, we fix without loss of generality the initial state to be fully polarized on ($Z_i=1$ for all $i$). We consider system sizes in the range $L=8\div 128$.

\bibliographystyle{apsrev4-2}

\end{document}